\documentclass[12pt]{article}

\author{Yu.~M.~Zinoviev
       \thanks{E-mail address: YURII.ZINOVIEV@IHEP.RU} \\
        {\it Institute for High Energy Physics} \\
        {\it Protvino, Moscow Region, 142280, Russia}}
\title{Massive supermultiplets with spin 3/2}

\date{}

\begin{document}

\maketitle

\begin{abstract}
In this paper we construct massive supermultiplets out of appropriate
set of massless ones in the same way as massive spin $s$ particle
could be constructed out of massless spin $s,s-1,\dots$ ones leading
to gauge invariant description of massive particle. Mainly we
consider massive spin 3/2 supermultiplets in a flat $d=4$ Minkowski
space both without central charge for $N=1,2,3$ as well as with
central charge for $N=2,4$. Besides, we give two examples of massive
$N=1$ supermultiplets with spin 3/2 and 2 in $AdS_4$ space.
\end{abstract}

\thispagestyle{empty}
\newpage
\setcounter{page}{1}

\section*{Introduction}

In a flat space-time massive spin $s$ particles in a massless limit
decompose into massless spin $s$, $s-1$, $\dots$ ones. This,
in particular, leads to the possibility of gauge invariant
description of massive spin $s$ particles, where massless spin $s$
field plays the role of ``main'' gauge field, while the lower spin
fields play the roles of Goldstone fields that have to be ``eaten``
in the process of spontaneous symmetry breaking to make main field
massive. Such approach to description of massive particles became
rather popular last times, e.g. \cite{Zin83}-\cite{BKR07}.

In the supersymmetric theories all particles must belong to some
supermultiplet, massive or massless. The same reasoning on the
massless limit means that massive supermultiplets could (should) be
constructed out of the massless ones in the same way as massive
particles out of the massless ones. Because supersymmetry is a very
restrictive symmetry even construction of free massive
supermultiplets could give very usefull and important information on
the structure of full interacting theories, where spontaneous
(super)symmetry breaking leading to the appearance of such massive
supermultiplets could occur. 

In supergravities partial super-Higgs effect $N \rightarrow N-k$, when
part of the supersymmetries remains unbroken, must unavoidably leads
to the appearance of $k$ massive spin 3/2 supermultiplets,
corresponding to the unbroken $N-k$ supersymmetries. The main subject
of our paper is the construction of massive spin 3/2 supermultiplets
out of the massless ones. Namely, we consider $N=1,2,3$
supermultiplets without central charge, as well as $N=2,4$
supermultiplets with central charge (for classification of massless
and massive supermultiplets see e.g. \cite{AAFL02}). We will not
consider one more possible case --- $N=6$ supermultiplet with central
charge, because it could hardly have phenomenological interest,
though it could be constructed in the same way as well. Besides, we
consider two examples of massive supermultiplets in (A)dS space,
namely $N=1$ spin 3/2 and spin 2 ones.

The paper is organized as follows. In the next section we start with
the simplest case --- massive $N=1$ spin 3/2 supermultiplet
\cite{FN83,Zin86,BGLP02,BGKP05}. This supermultiplets is known for a
long time, but it is very usefull to display the general technics for
construction of massive supermultiplets we will heavily use in what
follows. There is no strict definition of what is mass in (Anti) de
Sitter space and indeed rather different definitions there exist in
the literature. So we add small section devoted to the discussion of
this subject (and in particular the so called forbidden mass regions,
see e.g. \cite{DW01,DW01a,DW01c}) using massive spin 3/2 particle in
AdS space as an example. Then, in the next two sections, we consider
massive spin 3/2 \cite{AB99} and massive spin 2 supermultiplets
\cite{Zin02,BGPL02,BGKP05,GKTM06} in AdS space. Our results here show
rather interesting differences between supermultiplets in flat and AdS
spaces, as well as between supermultiplets with integer and
half-integer superspins. Also, massive spin 2 supermultiplet in AdS
shows one more example of the flat space limit -- massless limit
ambiguity, which is well known for the massive spin 2
\cite{KMP00,Por00,DDGV02} and spin 3/2 \cite{GN00,DW00,DLS02}
particles.

Then we return back to the flat Minkowski space and in the following
four sections we systematically consider massive spin 3/2
supermultiplets with $N=2$ and $N=3$ supersymmetry without central
charge as well as $N=2$ and $N=4$ supermultiplets with central charge.
In all cases exactly as in $N=1$ case it turns out crucial for the
whole construction to make duality transformations mixing different
supermultiplets containing vector fields.

\section{$N=1$ supermultiplet in flat space}

Let us start with the simplest case --- massive $N=1$ supermultiplet
\cite{FN83,Zin86,BGLP02,BGKP05} in flat space-time. Such multiplet
contains massive particles with spins $(3/2,1,1',1/2)$, all with equal
masses. In the massless limit massive spin 3/2 particle decompose into
massless spin 3/2 and 1/2 ones in the same way as massive spin 1
particle into massless spin1 and spin 0 ones. As a result in the
massless limit our massive supermultiplet gives three massless
supermultiplets:
$$
\left( \begin{array}{ccc}  & 3/2 & \\ 1 & & 1' \\ & 1/2 & \end{array}
\right) \quad \Rightarrow \quad \left( \begin{array}{c} 3/2 \\ 1
\end{array} \right) \oplus \left( \begin{array}{c} 1' \\
1/2 \end{array} \right) \oplus \left( \begin{array}{c}
1/2 \\ 0, 0' \end{array} \right)
$$
We denote appropriate fields as $(\Psi_\mu, A_\mu)$, $(B_\mu, \rho)$
and $(\chi, \varphi, \pi)$, correspondingly. We start with the
massless Lagrangian being the sum of kinetic terms for all these
fields:
\begin{equation}
{\cal L}_0 = \frac{i}{2} \varepsilon^{\mu\nu\alpha\beta}
\bar{\Psi}_\mu \gamma_5 \gamma_\nu \partial_\alpha \Psi_\beta +
\frac{i}{2} \bar{\rho} \hat{\partial} \rho + \frac{i}{2}
\bar{\chi} \hat{\partial} \chi - \frac{1}{4} A_{\mu\nu}{}^2 -
\frac{1}{4} B_{\mu\nu}{}^2 + \frac{1}{2} (\partial_\mu \varphi)^2 +
\frac{1}{2} (\partial_\mu \pi)^2
\end{equation}
which is invariant under three local gauge transformations:
$$
\delta \Psi_\mu = \partial_\mu \xi, \qquad \delta A_\mu =
\partial_\mu \lambda, \qquad \delta B_\mu = \partial_\mu
\tilde{\lambda}
$$
It is very important that the massive supermultiplet must contains
vector and axial-vector particles and not two vector or two
axial-vector ones. This, in turn, opens the possibility to make dual
transformation mixing two supermultiplet, namely $(\Psi_\mu, A_\mu)$
and $(B_\mu, \rho)$ ones. Thus, the most general supertransformations
leaving the massless Lagrangian invariant have the form:
\begin{eqnarray}
\delta \Psi_\mu &=& - \frac{i}{2\sqrt{2}} \sigma^{\alpha\beta} [
\cos(\theta) A_{\alpha\beta} - \sin(\theta) B_{\alpha\beta} \gamma_5
] \gamma_\mu \eta \nonumber \\
\delta A_\mu &=& \sqrt{2} \cos(\theta) (\bar{\Psi}_\mu \eta) + i
\sin(\theta) (\bar{\rho} \gamma_\mu \eta) \nonumber \\
\delta B_\mu &=& \sqrt{2} \sin(\theta) (\bar{\Psi}_\mu \gamma_5 \eta)
+ i \cos(\theta) (\bar{\rho} \gamma_\mu \gamma_5 \eta) \\
\delta \rho &=& - \frac{1}{2} \sigma^{\alpha\beta} [ \sin(\theta)
A_{\alpha\beta} + \cos(\theta) B_{\alpha\beta} \gamma_5 ] \eta
\nonumber \\
\delta \chi &=& - i \hat{\partial} (\varphi + \gamma_5 \pi) \eta
\quad \delta \varphi = (\bar{\chi} \eta) \quad \delta \pi =
(\bar{\chi} \gamma_5 \eta) \nonumber
\end{eqnarray}
Now we have to add mass terms for all fields as well as appropriate
corrections for the fermionic supertransformations. For this purpose
we, first of all, must identify Goldstone fields which have to be
eaten by gauge fields making them massive. For bosonic fields the
choice is unambiguous --- scalar field $\varphi$ for vector $A_\mu$
and pseudo-scalar $\pi$ for axial-vector $B_\mu$. Thus, we add the
following mass terms:
\begin{equation}
{\cal L}_m = - m A^\mu \partial_\mu \varphi - m B^\mu \partial_\mu
\pi + \frac{m^2}{2} A_\mu{}^2 + \frac{m^2}{2} B_\mu{}^2
\end{equation}
As for the spin 3/2 particle $\Psi_\mu$, we have two spinor fields
$\rho$ and $\chi$ which could serve as a Goldstone one, so we
consider the most general possible mass terms:
\begin{equation}
\frac{1}{m} {\cal L}_m = \frac{1}{2} \bar{\Psi}_\mu \sigma^{\mu\nu}
\Psi_\nu + i a_1 (\bar{\Psi} \gamma) \rho + i a_2 (\bar{\Psi}
\gamma) \chi + a_3 \bar{\rho} \rho + a_4 \bar{\rho} \chi + a_5
\bar{\chi} \chi
\end{equation}
Then the requirement that the total Lagrangian be invariant under
(corrected) supertransformations fixes the mixing angle $\theta$ as
well as all unknown coefficients:
$$
\sin(\theta) = \cos(\theta) = \frac{1}{\sqrt{2}}, \quad
a_1 = - \frac{1}{\sqrt{2}}, \quad a_2 = 1, \quad a_3 = 0, \quad a_4 =
- \sqrt{2}, \quad a_5 = \frac{1}{2}
$$
Moreover, this requirement unambiguously fixes the structure of
appropriate corrections for fermionic supertransformations:
\begin{eqnarray}
\frac{1}{m} \delta \Psi_\mu &=& [ A_\mu + B_\mu \gamma_5 - \frac{i}{2}
\gamma_\mu (\varphi + \gamma_5 \pi) ] \eta \nonumber \\
\frac{1}{m} \delta \rho &=&  - \frac{1}{\sqrt{2}} ( \varphi +
 \gamma_5 \pi ) \eta \\
\frac{1}{m} \delta \chi &=& [ i \hat{A} + i \hat{B} \gamma_5 +
\varphi + \gamma_5 \pi ] \eta  \nonumber
\end{eqnarray}
It is easy to check that with the resulting fermionic mass terms:
$$
\frac{1}{m} {\cal L}_m = \frac{1}{2} \bar{\Psi}_\mu \sigma^{\mu\nu}
\Psi_\nu - \frac{i}{\sqrt{2}} (\bar{\Psi} \gamma) \rho + i
(\bar{\Psi} \gamma) \chi - \sqrt{2} \bar{\rho} \chi + \frac{1}{2}
\bar{\chi} \chi
$$
the total Lagrangian is invariant (besides global
supertransformations) under the following local gauge transformations:
\begin{equation}
\delta \Psi_\mu = \partial_\mu \xi - \frac{im}{2} \gamma_\mu \xi,
\qquad \delta \rho = - \frac{m}{\sqrt{2}} \xi, \qquad
\delta \chi = m \xi
\end{equation}
From the last formula one can see which combination of two spinor
fields plays the role of Goldstone one. Indeed, if we introduce two
orthogonal combinations:
$$
\tilde{\rho} = - \frac{1}{\sqrt{3}} \rho + \sqrt{\frac{2}{3}} \chi,
\qquad \tilde{\chi} = \sqrt{\frac{2}{3}} \rho + \frac{1}{\sqrt{3}}
\chi
$$
then the fermionic mass terms take the form:
$$
\frac{1}{m} {\cal L}_m = \frac{1}{2} \bar{\Psi}_\mu \sigma^{\mu\nu}
\Psi_\nu + i \sqrt{\frac{3}{2}} (\bar{\Psi} \gamma) \tilde{\rho} +
\tilde{\bar{\rho}} \tilde{\rho} - \frac{1}{2} \tilde{\bar{\chi}}
\tilde{\chi}
$$
which explicitly shows that we have spin 3/2 and spin 1/2 particles
with equal masses. Moreover, by using this local gauge transformation
with $\xi = - (\varphi + \gamma_5 \pi) \eta$ and introducing gauge
invariant derivatives for the scalar fields:
$$
\nabla_\mu \varphi = \partial_\mu \varphi - m A_\mu, \qquad
\nabla_\mu \pi = \partial_\mu \pi - m B_\mu
$$
one can bring supertransformations for the fermions into the following
simple form:
\begin{eqnarray}
\delta \Psi_\mu &=& [ - \frac{i}{4} \sigma^{\alpha\beta} (
A_{\alpha\beta} -  B_{\alpha\beta} \gamma_5 ) \gamma_\mu - \nabla_\mu
(\varphi + \gamma_5 \pi) ] \eta \nonumber  \\
\delta \rho &=& - \frac{1}{2\sqrt{2}} \sigma^{\alpha\beta} [
A_{\alpha\beta} + B_{\alpha\beta} \gamma_5 ] \eta \qquad
\delta \chi = - i \hat{\nabla} (\varphi + \gamma_5 \pi) \eta
\end{eqnarray}

Note here that we work with Majorana fermions (and Majorana
representation of $\gamma$-matrices). In this, the $\gamma_5$ matrix
plays the role of imaginary unit $i$. Then we can further simplify
formula given above by introducing complex objects $C_\mu = (A_\mu +
\gamma_5 B_\mu)$ and $z = \varphi + \gamma_5 \pi$:
\begin{eqnarray}
\delta \Psi_\mu &=& [ - \frac{i}{4} \sigma^{\alpha\beta} 
\bar{C}_{\alpha\beta} \gamma_\mu - \nabla_\mu z ] \eta \nonumber  \\
\delta \rho &=& - \frac{1}{2\sqrt{2}} \sigma^{\alpha\beta} 
C_{\alpha\beta} \eta \qquad \delta \chi = - i \hat{\nabla} z \eta 
\end{eqnarray}
Now it is evident that we have one more symmetry --- axial $U(1)_A$
global symmetry, the axial charges for all fields being:
\begin{center}
\begin{tabular}{|c|c|c|c|} \hline
field  & $\eta$ & $\Psi_\mu$, $\rho$, $\chi$ & $C_\mu$, $z$ \\
\hline $q_A$ & +1 & 0 & --1 \\ \hline
\end{tabular}
\end{center}

Thus, we have seen that it is important for construction of this
supermultiplet to have possibility of making dual rotation of 
vector fields mixing massless supermultiplets. Also, there is a tight
connection between vector fields (Higgs effect) and spin 3/2
(super-Higgs effect) masses. And indeed, the existence of dual
versions of $N=2$ supergravities and appropriate gaugings makes
partial super-Higgs effect possible
\cite{CGP86a,Zin86,Zin92,Zin94,AAFL02c}.

\section{Massive spin 3/2 in $(A)dS_4$}

In the (Anti) de Sitter space-time it is the (Anti) de Sitter group
that plays the role of global background symmetry instead of Poincare
group in Minkowski space. As a result, there is no strict definition
of what is mass in such space. And indeed, a lot of controversy on
this subject exists in the literature. The aim of this small section
is to explain the definition of mass we (personally) adhere to using
massive spin 3/2 particle as an example. 

Anti de Sitter space is the constant curvature space without torsion
or non-metricity, so the main difference from the Minkowski space is
the replacement of ordinary partial derivatives by the covariant
ones. We will use the following normalization here:
\begin{equation}
[ \nabla_\mu, \nabla_\nu ] = \frac{\kappa}{2} \sigma_{\mu\nu}, \qquad
\kappa = - \frac{2\Lambda}{(d-1)(d-2)} = - \frac{\Lambda}{3}
\end{equation}
where $\Lambda$ --- cosmological term. Now let us consider the
quadratic Lagrangian for spin 3/2 $\Psi_\mu$ and spin 1/2 $\chi$
fields with the most general mass terms:
\begin{equation}
{\cal L} = \frac{i}{2} \varepsilon^{\mu\nu\alpha\beta} \bar{\Psi}_\mu
\gamma_5 \gamma_\nu \nabla_\alpha \Psi_\beta + \frac{i}{2}
\bar{\chi} \hat{\nabla} \chi + \frac{M}{2} \bar{\Psi}_\mu
\sigma^{\mu\nu} \Psi_\nu + i a_1 (\bar{\Psi} \gamma) \chi +
\frac{a_2}{2} \bar{\chi} \chi
\end{equation}
and require that it will be invariant under the following local gauge
transformations:
$$
\delta \Psi_\mu = \nabla_\mu \xi + i \alpha_1 \gamma_\mu \xi \qquad
\delta \chi = \alpha_2 \xi
$$
Simple calculations immediately give:
$$
a_1 = \alpha_2, \qquad a_2 = 2 M, \qquad \alpha_1 = - \frac{M}{2},
\qquad M^2 = \frac{2}{3} \alpha_2{}^2 + k
$$
For the gauge invariant description of massive particles it is
natural to define massless limit as the limit when Goldstone field(s)
completely decouples from the the main gauge field. In the case at
hands this means that it is the parameter $a_1$ determines the mass
$a_1 \sim m$. As for the concrete normalization we will require that
in the flat space limit our definition coincides with the usual one.
Thus $a_1 = \sqrt{\frac{3}{2}} m$ and $M = \sqrt{m^2 + \kappa}$. One
of the peculiar features of (Anti) de Sitter spaces is the existence
of so called forbidden mass regions \cite{DW01,DW01a,DW01c}. And we
see that different choices of what one call mass ($M$ or $m$ in the
case considered) lead to drastically different physical
interpretations, as we illustrate by the following simple picture:
\begin{center}
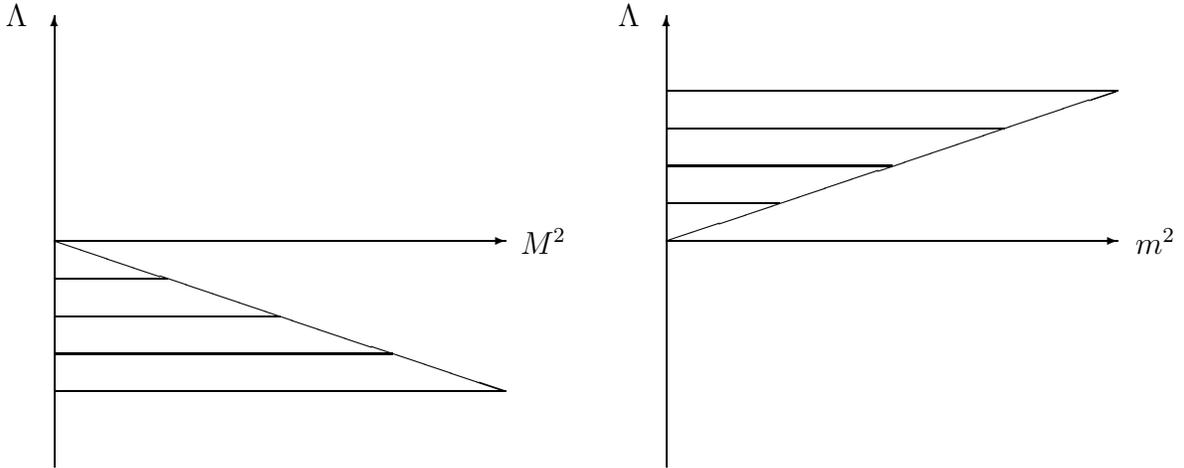
\begin{figure}[htb]
\setlength{\unitlength}{1mm}
\noindent
\begin{picture}(80,80)
\put(10,40){\vector(1,0){60}}
\put(10,10){\vector(0,1){60}}
\put(0,65){\makebox(10,10)[]{$\Lambda$}}
\put(70,35){\makebox(10,10)[]{$M^2$}}
\put(10,40){\line(3,-1){60}}
\put(10,35){\line(1,0){15}}
\put(10,30){\line(1,0){30}}
\put(10,25){\line(1,0){45}}
\put(10,20){\line(1,0){60}}
\end{picture}
\begin{picture}(80,80)
\put(10,40){\vector(1,0){60}}
\put(10,10){\vector(0,1){60}}
\put(0,65){\makebox(10,10)[]{$\Lambda$}}
\put(70,35){\makebox(10,10)[]{$m^2$}}
\put(10,40){\line(3,1){60}}
\put(10,45){\line(1,0){15}}
\put(10,50){\line(1,0){30}}
\put(10,55){\line(1,0){45}}
\put(10,60){\line(1,0){60}}
\end{picture}
\caption{Forbidden regions.}
\end{figure}
\end{center}
For the fermionic fields with higher spins similar results can be
easily obtained from that of \cite{Met06}. Also note the paper
\cite{Gar03} where group-theoretical arguments in favor of de Sitter
space were given. At the same time for the bosonic particles our
definition agrees perfectly with the one used by authors of
\cite{DW01,DW01a,DW01c}.

\section{$N=1$ supermultiplet in $AdS_4$}

In this section we consider the same massive $N=1$ supermultiplet in
Anti de Sitter space \cite{AB99}. Now, besides the replacement of
ordinary partial derivatives by the covariant ones, one has to take
care on the definition of global supertransformations. The simple and
natural choice (e.g. \cite{IS80}) is to use the spinor $\eta$
satisfying the relation:
$$
\nabla_\mu \eta = - \frac{i\kappa_0}{2} \gamma_\mu \eta, \qquad
\kappa_0{}^2 = \kappa
$$
as a parameter of such ``global'' supertransformations.

Now we return back to the sum of kinetic terms for all fields where
ordinary derivatives are replaced by covariant ones:
\begin{equation}
{\cal L}_0 = \frac{i}{2} \varepsilon^{\mu\nu\alpha\beta}
\bar{\Psi}_\mu \gamma_5 \gamma_\nu \nabla_\alpha \Psi_\beta +
\frac{i}{2} \bar{\rho} \hat{\nabla} \rho + \frac{i}{2}
\bar{\chi} \hat{\nabla} \chi - \frac{1}{4} A_{\mu\nu}{}^2 -
\frac{1}{4} B_{\mu\nu}{}^2 + \frac{1}{2} (\partial_\mu \varphi)^2 +
\frac{1}{2} (\partial_\mu \pi)^2
\end{equation}
In Anti de Sitter space this Lagrangian is no longer invariant under
the initial supertransformations:
$$
\delta_0 {\cal L}_0 = - \frac{i\kappa_0}{\sqrt{2}} \bar{\Psi}_\mu [
\cos(\theta) (A^{\mu\nu} - \gamma_5 \tilde{A}^{\mu\nu}) +
\sin(\theta) (\gamma_5 B^{\mu\nu} - \tilde{B}^{\mu\nu}) ] \gamma_\nu
\eta + i\kappa_0 \bar{\chi} \gamma^\mu (\partial_\mu \varphi -
\gamma_5 \partial_\mu \pi) \eta
$$
We proceed by adding the most general mass terms for the fermions as
well as one derivative terms for the bosons:
\begin{eqnarray}
{\cal L}_1 &=& \frac{a_1}{2} \bar{\Psi}_\mu \sigma^{\mu\nu} \Psi_\nu +
i a_2 (\bar{\Psi} \gamma) \rho + i a_3 (\bar{\Psi} \gamma) \chi +
a_4 \bar{\rho} \rho + a_5 \bar{\rho} \chi + \frac{a_6}{2}
\bar{\chi} \chi - \nonumber \\
 && - m_1 A^\mu \partial_\mu \varphi - m_2 B^\mu \partial_\mu \pi
\end{eqnarray}
and the most general additional terms for the fermionic
supertransformations:
\begin{eqnarray}
\delta_1 \Psi_\mu &=& [ \alpha_1 A_\mu + \alpha_2 B_\mu \gamma_5
+ i \alpha_3 \gamma_\mu \varphi + i \alpha_4 \gamma_\mu \gamma_5 \pi ]
\eta \nonumber \\
\delta_1 \rho &=& [ i \beta_1 \hat{A} + i \beta_2 \hat{B} \gamma_5 +
\beta_3 \varphi + \beta_4 \gamma_5 \pi ] \eta \\
\delta_1 \chi &=& [ i \beta_5 \hat{A} + i \beta_6 \hat{B} \gamma_5 +
\beta_7 \varphi + \beta_8 \gamma_5 \pi ] \eta \nonumber
\end{eqnarray}
Requirement that all variations containing one derivative cancel
gives:
$$
a_1 = - M, \quad a_2 = \frac{M}{\sqrt{2}} \sin(2\theta), \quad 
\alpha_1 = - M \sqrt{2} \cos(\theta) \quad
\alpha_2 = - M \sqrt{2} \sin(\theta)
$$
$$
a_3 = m_1 \sqrt{2} \cos(\theta) = m_2 \sqrt{2} \sin(\theta), \qquad 
\alpha_3 = \alpha_4 = - \frac{a_3}{2}
$$
$$
a_4 = \beta_1 = \beta_2 = 0, \qquad
\beta_3 = a_5 + m_1 \sin(\theta), \qquad
\beta_4 = a_5 + m_2 \cos(\theta)
$$
$$
a_5 = - \frac{m_1}{\sin(\theta)} = -
\frac{m_2}{\cos(\theta)}, \quad
\beta_5 = m_1, \quad \beta_6 = m_2, \quad
\beta_7 = a_6 - \kappa_0, \quad \beta_8 = a_6 + \kappa_0
$$
Here $M = \frac{\kappa_0}{\cos(2\theta)}$. We see that it is the
mixing angle $\theta$ (together with cosmological term) determines
all masses in this case. Recall, that in flat space we have
$\sin(\theta) = \cos(\theta) = \frac{1}{\sqrt{2}}$, while here it is
the singular point. Indeed, the flat space results could be correctly
reproduced only by taking simultaneous limits $\kappa_0 \rightarrow 0$
and $\theta \rightarrow \frac{\pi}{4}$ so that $\kappa_0
\tan(2\theta)$ remains to be fixed.

At last we add appropriate mass terms for bosons: 
\begin{equation}
{\cal L}_2 = \frac{m_1{}^2}{2} A_\mu{}^2 + \frac{m_2{}^2}{2}
B_\mu{}^2 + b_1 \varphi^2 + b_2 \pi^2
\end{equation}
and require that all variations without derivatives cancel. This
gives:
$$
m_1 = M \sqrt{2} \sin(\theta), \qquad m_2 = M \sqrt{2} \cos(\theta),
\qquad a_6 = - M, \qquad b_1 = b_2 = 0
$$

The resulting mass terms for the fermions look like:
\begin{equation}
{\cal L}_m = - \frac{M}{2} \bar{\Psi}_\mu \sigma^{\mu\nu} \Psi_\nu +
\frac{i M \sin(2\theta)}{\sqrt{2}} (\bar{\Psi} \gamma) \rho + i M
\sin(2\theta) (\bar{\Psi} \gamma) \chi - M \sqrt{2} \bar{\rho} \chi
- \frac{M}{2} \bar{\chi} \chi \\
\end{equation}
In this, besides the supertransformations, Lagrangian is invariant
under the following local gauge transformations:
\begin{equation}
\delta \Psi_\mu = \nabla_\mu \xi + \frac{i M}{2} \gamma_\mu \xi,
\qquad \delta \rho = \frac{M \sin(2\theta)}{\sqrt{2}} \xi, \qquad
\delta \chi = M \sin(2\theta) \xi
\end{equation}
Comparing this formula with the results of previous sections, one can
conclude that it is the combination $m = M \sin(2\theta) = \kappa_0
\tan(2\theta)$ determines the mass for spin 3/2 particle. So we have
four massive fields with masses (which become equal in the limit
$\theta \rightarrow \pi/4$):
\begin{equation}
m_{3/2} = m, \quad m_1 = \frac{m}{\sqrt{2} \cos(\theta)},
\quad m_{1'} = \frac{m}{\sqrt{2}\sin(\theta)}, \quad
m_{1/2} = \frac{m}{\sin(2\theta)}
\end{equation}

As in the flat case, introducing gauge invariant derivatives for
scalar fields:
$$
\nabla_\mu \varphi = \partial_\mu \varphi - m_1 A_\mu, \qquad
\nabla_\mu \pi = \partial_\mu \pi - m_2 B_\mu, \qquad
$$
and making local gauge transformation with
$ \xi = (\cot(\theta) \varphi + \tan(\theta) \pi) \eta $
one can bring supertransformations for fermions into relatively
simple form:
\begin{eqnarray}
\delta \Psi_\mu &=& - \frac{i}{2\sqrt{2}} \sigma^{\alpha\beta} [
\cos(\theta) A_{\alpha\beta} - \sin(\theta) B_{\alpha\beta} \gamma_5
] \gamma_\mu \eta + ( \cot(\theta) \nabla_\mu \varphi + \tan(\theta) 
\nabla_\mu \pi \gamma_5) \eta \nonumber \\
\delta \rho &=& - \frac{1}{2} \sigma^{\alpha\beta} [ \sin(\theta)
A_{\alpha\beta} + \cos(\theta) B_{\alpha\beta} \gamma_5 ] \eta \qquad
\delta \chi = - i \hat{\nabla} (\varphi + \gamma_5 \pi) \eta
\end{eqnarray}
Note, that in Anti de Sitter space there is no axial $U(1)_A$
symmetry.

\section{$S=2$ supermultiplet in $AdS_4$}

This paper devoted mainly to construction of massive spin 3/2
supermultiplets, but it is instructive to compare with the next to
simplest case --- massive spin 2 supermultiplet. In flat space such
multiplets were constructed in \cite{Zin02} (see also
\cite{BGPL02,GSS04,GKTM06}), so we consider $AdS_4$ case. Massive
$N=1$ spin 2 supermultiplet contains four massive fields
$(2,3/2,3/2',1)$. Taking into account that in the massless limit (in
flat space, see below) massive spin 2 particle decompose into massless
spin 2, spin 1 and spin 0 ones, we have to use four massless
supermultiplets for our construction:
$$
\left( \begin{array}{ccc}  & 2 & \\ \frac{3}{2} & & \frac{3}{2}' \\ &
1 & \end{array}
\right) \quad \Rightarrow \quad 
\left( \begin{array}{c} 2 \\ \frac{3}{2} \end{array} \right) \oplus
\left( \begin{array}{c} \frac{3}{2}' \\ 1 \end{array} \right) \oplus
\left( \begin{array}{c} 1' \\ \frac{1}{2} \end{array} \right) \oplus
\left( \begin{array}{c} \frac{1}{2}' \\ 0,0' \end{array} \right)
$$

We denote appropriate fields as $(h_{\mu\nu},\Psi_\mu)$,
$(\Omega_\mu,A_\mu)$, $(B_\mu,\rho)$ and $(\chi,\varphi,\pi)$ and
start with the sum of kinetic terms for all fields (with ordinary
partial derivatives replaced by covariant ones):
\begin{eqnarray}
{\cal L}_0 &=& \frac{1}{2} \nabla^\mu h^{\alpha\beta} \nabla_\mu
h_{\alpha\beta} - (\nabla h)^\mu (\nabla h)_\mu + (\nabla h)^\mu
\nabla_\mu h - \frac{1}{2} \nabla^\mu h \nabla_\mu h - \nonumber \\
 && - \frac{1}{4} A_{\mu\nu}{}^2 + \frac{1}{2} (\partial_\mu
\varphi)^2 - \frac{1}{4} B_{\mu\nu}{}^2 + \frac{1}{2} (\partial_\mu
\pi)^2 + \\
 && + \frac{i}{2} \varepsilon^{\mu\nu\alpha\beta} \bar{\Psi}_\mu
\gamma_5 \gamma_\nu \nabla_\alpha \Psi_\beta + \frac{i}{2}
\varepsilon^{\mu\nu\alpha\beta} \bar{\Omega}_\mu \gamma_5 \gamma_\nu
\nabla_\alpha \Omega_\beta + \frac{i}{2} \bar{\rho} \hat{\nabla}
\rho + \frac{i}{2} \bar{\chi} \hat{\nabla} \chi \nonumber
\end{eqnarray}
It is crucial for the whole construction that we again have one
vector and one axial-vector fields and the possibility to make dual
mixing of two supermultiplets containing these fields. So we will use
the following ansatz for supertransformations:
\begin{eqnarray}
\delta_0 h_{\mu\nu} &=& i ( \bar{\Psi}_{(\mu} \gamma_{\nu)} \eta)
\qquad \delta_0 \Psi_\mu = - \sigma^{\alpha\beta} \nabla_\alpha
h_{\beta\mu} \nonumber \\
\delta_0 \Omega_\mu &=& - \frac{i}{2\sqrt{2}} \sigma^{\alpha\beta}
(\cos(\theta) A_{\alpha\beta} - \sin(\theta) B_{\alpha\beta}
\gamma_5) \gamma_\mu \eta \nonumber \\
\delta_0 A_\mu &=& \sqrt{2} \cos(\theta) (\bar{\Omega}_\mu \eta) + i
\sin(\theta) (\bar{\rho} \gamma_\mu \eta) \\
\delta_0 B_\mu &=& \sqrt{2} \sin(\theta) (\bar{\Omega}_\mu \gamma_5
\eta) + i \cos(\theta) (\bar{\rho} \gamma_\mu \gamma_5 \eta)
\nonumber \\
\delta_0 \rho &=& - \frac{1}{2} \sigma^{\alpha\beta} (\sin(\theta)
A_{\alpha\beta} + \cos(\theta) B_{\alpha\beta} \gamma_5) \eta 
\nonumber \\
\delta_0 \chi &=& - i \gamma_\mu (\partial_\mu \varphi + \partial_\mu
\pi \gamma_5) \eta \qquad \delta_0 \varphi = (\bar{\chi} \eta) \qquad
\delta_0 \pi = (\bar{\chi} \gamma_5 \eta) \nonumber
\end{eqnarray}
In AdS space the sum of kinetic terms is not invariant under these
transformations any more and we must take it into account in the
subsequent calculations. The next question is which fields play the
role of Goldstone ones making gauge fields massive. The choice for
bosonic fields is unambiguous --- vector $A_\mu$ and scalar $\varphi$
fields for $h_{\mu\nu}$ and pseudo-scalar $\pi$ for $B_\mu$ one. But
for the fermions situation is more complicated. Recall that in AdS
case we have no axial $U(1)_A$ symmetry which could restrict possible
choice, thus we have to consider the most general mass terms for the
fermions. So we add to our Lagrangian:
\begin{eqnarray}
{\cal L}_1 &=& m \sqrt{2} (h^{\mu\nu} \nabla_\mu A_\nu - h (\nabla
A)) - M \sqrt{3} A^\mu \partial_\mu \varphi - \tilde{m} B^\mu
\partial_\mu \pi - \nonumber \\
 && - \frac{a_1}{2} \bar{\Psi}_\mu \sigma^{\mu\nu} \Psi_\nu - a_2
\bar{\Psi}_\mu \sigma^{\mu\nu} \Omega_\nu - \frac{a_3}{2}
\bar{\Omega}_\mu \sigma^{\mu\nu} \Omega_\nu + i a_4 (\bar{\Psi}
\gamma) \rho + i a_5 (\bar{\Psi} \gamma) \chi + \nonumber \\
 && + i a_6 (\bar{\Omega} \gamma) \rho + i a_7 (\bar{\Omega} \gamma)
\chi + \frac{a_8}{2} \bar{\rho} \rho + a_9 \bar{\rho} \chi +
\frac{a_{10}}{2} \bar{\chi} \chi
\end{eqnarray}
where $M = \sqrt{m^2 + 2\kappa}$, and require that all variations
with one derivative cancel (making necessary corrections for
fermionic supertransformations). This gives us:
$$
\sin(\theta) = \frac{\sqrt{3}}{2}, \qquad \cos(\theta) = \frac{1}{2},
\qquad \tilde{m} = M
$$
$$
a_1 = \kappa_0, \quad a_2 = m, \quad a_3 = - 2\kappa_0, \quad a_4 = m
\sqrt{\frac{3}{2}}, \quad a_5 = 0
$$
$$
a_6 = - \sqrt{\frac{3}{2}} \kappa_0, \quad a_7 = \sqrt{\frac{3}{2}}
M, \quad a_8 = 0, \quad a_9 = - 2 M
$$
Note that in sharp contrast with the massive spin 3/2 case now the
mixing angle $\theta$ is fixed (and has the same value as in flat
case) so all masses are determined by spin 2 mass $m$ and
cosmological constant $\kappa$. We proceed by adding appropriate mass
terms for bosonic fields:
\begin{equation}
{\cal L}_2 = - \frac{m^2 - 2\kappa}{2} h^{\mu\nu} h_{\mu\nu} +
\frac{m^2 + \kappa}{2} h^2 - \sqrt{\frac{3}{2}} m M h \varphi + 3
\kappa A_\mu{}^2 + m^2 \varphi^2 + \frac{M^2}{2} B_\mu{}^2
\end{equation}
and requiring cancellation of all variations without derivatives. This
fixes the last unknown parameter $a_{10} = 2 \kappa_0$ and the
structure of additional terms in fermionic supertransformations:
\begin{eqnarray}
\delta_1 \Psi_\mu &=& [ i \kappa_0 h_{\mu\nu} \gamma^\nu -
\frac{m}{\sqrt{2}} \gamma_\mu \hat{A} - m \sqrt{\frac{3}{2}} \gamma_5
B_\mu ] \eta \nonumber \\
\delta_1 \Omega_\mu &=& [ i m h_{\mu\nu} \gamma^\nu + \kappa_0
\sqrt{2} A_\mu + \kappa_0 \sqrt{6}  \gamma_5 B_\mu - \frac{i}{2}
\sqrt{\frac{3}{2}} M \gamma_\mu (\varphi + \gamma_5 \pi) ] \eta 
\nonumber  \\
\delta_1 \rho &=& [ - \frac{M}{2} \varphi - \frac{3M}{2} \gamma_5
\pi ] \eta \\
\delta_1 \chi &=& [ i M \sqrt{3} \hat{A} + i M \hat{B} \gamma_5
+ \kappa_0 \varphi + 3\kappa_0 \gamma_5 \pi ] \eta \nonumber
\end{eqnarray}
Recall that in the flat case \cite{Zin02} due to axial $U(1)_A$
symmetry fermionic mass terms were the Dirac ones. For the non-zero
cosmological term the structure of these terms become more
complicated:
\begin{eqnarray}
{\cal L}_m &=& - \frac{\kappa_0}{2} \bar{\Psi}_\mu \sigma^{\mu\nu}
\Psi_\nu - m \bar{\Psi}_\mu \sigma^{\mu\nu} \Omega_\nu - \kappa_0
\bar{\Omega}_\mu \sigma^{\mu\nu} \Omega_\nu + i m \sqrt{\frac{3}{2}}
(\bar{\Psi} \gamma) \rho - \nonumber \\
 && - i \kappa_0 \sqrt{\frac{3}{2}} (\bar{\Omega} \gamma) \rho + i M
\sqrt{\frac{3}{2}} (\bar{\Omega} \gamma) \chi - 2 M \bar{\rho} \chi
+ \kappa_0 \bar{\chi} \chi
\end{eqnarray}
Nevertheless, it is not hard to check that the Lagrangian obtained is
invariant (besides supertransformations) under two local gauge
transformations:
\begin{eqnarray}
\delta \Psi_\mu &=& \nabla_\mu \xi_1 + \frac{i\kappa_0}{2} \gamma_\mu
\xi_1, \qquad \delta \Omega_\mu = \frac{im}{2} \gamma_\mu \xi_1,
\qquad \delta \rho = m \sqrt{\frac{3}{2}} \xi_1  \\
\delta \Psi_\mu &=& \frac{im}{2} \gamma_\mu \xi_2, \qquad \delta
\Omega_\mu = \nabla_\mu \xi_2 - i \kappa_0 \gamma_\mu \xi_2, \qquad
\delta \rho = - \kappa_0 \sqrt{\frac{3}{2}} \xi_2, \qquad \delta
\chi = M \sqrt{\frac{3}{2}} \xi_2 \nonumber
\end{eqnarray}

As is known \cite{Zin01,BHR05}, in the (A)dS space massive spin s
particle decompose in the massless limit into massless spin s and
massive spin s-1 ones. Similarly, in the massless limit $m \rightarrow
0$ our massive spin 2 supermultiplets decompose into massless
$(2,3/2)$ supermultiplet and massive $(3/2,1,1',1/2)$  one with mass
$M = 2\kappa_0$ and mixing angle $\sin(2\theta) = \frac{\sqrt{3}}{2}$.
Note that in the paper \cite{AB99} this corresponds to the value
$\varepsilon = 1/2$.

All the formulas could be greatly simplified if we introduce gauge
invariant derivatives:
\begin{equation}
D_\mu h_{\alpha\beta} = \nabla_\mu h_{\alpha\beta} -
\frac{m}{\sqrt{2}} A_\mu g_{\alpha\beta}, \qquad
D_\mu \varphi = \partial_\mu \varphi - M \sqrt{3} A_\mu, \qquad
D_\mu \pi = \partial_\mu \pi - M B_\mu
\end{equation}
as well as notation $H_{\mu\nu} = h_{\mu\nu} - \frac{m}{M\sqrt{6}}
\varphi g_{\mu\nu}$ and make two local gauge transformations with the
parameters:
$$
\xi_1 = \frac{m}{M\sqrt{6}} (\varphi + 3 \pi \gamma_5) \eta \qquad
\xi_2 = - \frac{\kappa_0 \sqrt{6}}{M} (\frac{1}{3} \varphi + \pi
\gamma_5) \eta
$$
Then the resulting supertransformations for the fermions take the
form:
\begin{eqnarray}
\delta \Psi_\mu &=& [ - \sigma^{\alpha\beta} D_\alpha h_{\beta\mu} +
\frac{m}{M\sqrt{6}} (D_\mu \varphi + 3 D_\mu \pi \gamma_5) + i
\kappa_0 H_{\mu\nu} \gamma^\nu ] \eta \nonumber \\
\delta \Omega_\mu &=& [ - \frac{i}{4\sqrt{2}} \sigma^{\alpha\beta}
( A_{\alpha\beta} - \sqrt{3} B_{\alpha\beta} \gamma_5) \gamma_\mu
- \frac{\kappa_0 \sqrt{6}}{M} ( \frac{1}{3} D_\mu \varphi + D_\mu \pi
\gamma_5) + i m H_{\mu\nu} \gamma^\nu ] \eta \\
\delta \rho &=& - \frac{1}{4} \sigma^{\alpha\beta} (\sqrt{3}
A_{\alpha\beta} + B_{\alpha\beta} \gamma_5) \eta \qquad
\delta \chi = - i \gamma_\mu (D_\mu \varphi + D_\mu \pi \gamma_5)
\eta \nonumber
\end{eqnarray}
It is interesting that the structure of terms containing scalar
fields gives us one more example of flat space limit --- massless
limit ambiguity well known for the massive spin 2 
\cite{KMP00,Por00,DDGV02} and spin 3/2 \cite{GN00,DW00,DLS02}
particles. Indeed, if one takes massless limit keeping cosmological
term fixed, one gets:
$$
\delta \Psi_\mu \sim 0, \qquad
\delta \Omega_\mu \sim -\sqrt{3} ( \frac{1}{3} D_\mu \varphi + D_\mu
\pi \gamma_5) \eta
$$
At the same time, in the flat space limit with fixed $m$ we get:
$$
\delta \Psi_\mu \sim \frac{1}{\sqrt{6}} (D_\mu \varphi + 3 D_\mu \pi
\gamma_5) \eta \qquad \delta \Omega_\mu \sim 0
$$

\section{$N=2$ supermultiplet}

Now we return back to flat Minkowski space and consider massive spin
3/2 supermultiplets with extended supersymmetries. Our next example
--- massive $N=2$ supermultiplet containing one spin 3/2, four spin
1, six spin 1/2 and four spin 0 particles. Simple calculations show
that in the massless limit we obtain one spin 3/2 supermultiplet,
doublet of vector supermultiplets and one hypermultiplet:
$$
\left( \begin{array}{c} \frac{3}{2} \\ 4 \otimes 1 \\ 6
\otimes \frac{1}{2} \\ 4 \otimes 0 \end{array} \right) \quad
\Rightarrow \quad \left( \begin{array}{c} \frac{3}{2} \\
2 \otimes 1 \\ \frac{1}{2}\end{array} \right) \oplus
2 \otimes \left( \begin{array}{c}  1 \\ 2 \otimes \frac{1}{2}  \\
2 \otimes 0 \end{array} \right) \oplus \left(
\begin{array}{c} 2 \otimes \frac{1}{2} \\ 4 \otimes 0 \end{array}
\right)
$$
We denote all these fields as $(\Psi_\mu, A_\mu{}^i, \rho)$,
$(B_\mu{}^i, \Omega_i{}^j, z_i)$ and $(\chi, \lambda, \Phi_i)$ and
start with the sum of their kinetic terms:
\begin{eqnarray}
{\cal L}_0 &=& \frac{i}{2} \varepsilon^{\mu\nu\alpha\beta}
\bar{\Psi}_\mu \gamma_5 \gamma_\nu \partial_\alpha \Psi_\beta -
\frac{1}{4} A_{\mu\nu}{}^2 - \frac{1}{4} B_{\mu\nu}{}^2 + \frac{i}{2}
\bar{\rho} \hat{\partial} \rho + \frac{i}{2} \bar{\Omega}_i{}^j
\hat{\partial} \Omega_i{}^j + \nonumber \\
 && + \frac{i}{2} \bar{\chi} \hat{\partial} \chi + \frac{i}{2}
\bar{\lambda} \hat{\partial} \lambda + \frac{1}{2} \partial_\mu
\bar{z}^i \partial_\mu z_i + \frac{1}{2} \partial_\mu \bar{\Phi}^i
\partial_\mu \Phi_i
\end{eqnarray}
It is important that the massive supermultiplet we are going to
construct must have total $U(2) = SU(2) \otimes U(1)_A$ symmetry. It
is again crucial that we have doublet of vector and doublet of
axial-vector fields in our disposal. This allows us by making dual
transformation mixing two vector supermultiplets introduce complex
objects $C_\mu{}^i = A_\mu{}^i + \gamma_5 B_\mu{}^i$. Also, this
$U(2)$ symmetry dictates our choice of parametrisation for
hypermultiplet (there exists three different ones). Thus we take the
following form of supertransformations for massless supermultiplets:
\begin{eqnarray}
\delta \Psi_\mu &=& - \frac{i}{4} \sigma^{\alpha\beta} \gamma_\mu 
C_{\alpha\beta}{}^i \eta_i \qquad 
\delta \rho = - \frac{1}{2\sqrt{2}} \sigma^{\alpha\beta}
\bar{C}_{\alpha\beta i} \varepsilon^{ij} \eta_j \nonumber \\
\delta \bar{C}_{\mu i} &=& 2 ( \bar{\Psi}_\mu \eta_i) +
i\sqrt{2} ( \bar{\Omega}_j{}^i \gamma_\mu \eta_j) \qquad
\delta C_\mu{}^i = i\sqrt{2} ( \bar{\rho} \gamma_\mu
\varepsilon^{ij} \eta_j) \\
\delta \Omega_i{}^j &=& - \frac{1}{2\sqrt{2}} \sigma^{\alpha\beta}
C_{\alpha\beta}{}^j \eta_i - i
\varepsilon_{ik} \hat{\partial} z_j \eta^k 
\qquad   \delta \bar{z}^i = 2 \varepsilon^{jk}
(\bar{\Omega}_j{}^i \eta_k) \nonumber
\end{eqnarray}
\begin{eqnarray}
\delta \chi &=& i \hat{\partial} \varepsilon^{ij} \Phi_i \eta_j \qquad
\delta \lambda = - i \hat{\partial} \bar{\Phi}^i \eta_i \nonumber \\
\delta \bar{\Phi}^i &=& - 2 \varepsilon^{ij} (\bar{\chi} \eta_j)
\qquad \delta \Phi_i = 2 (\bar{\lambda} \eta_i) 
\end{eqnarray}
In the complex notations the $SU(2)$ symmetry of our construction is
explicit, while the axial $U(1)_A$ symmetry is achieved by the
following assignment of axial charges:
\begin{center}
\begin{tabular}{|c|c|c|c|c|c|} \hline
field & $\rho$ & $\eta_i$ & $\Psi_\mu$, $\Omega_i{}^j$, $\chi$ &
$C_\mu{}^i$, $z_i$, $\Phi_i$ & $\lambda$ \\ \hline
$q_A$ & +2 & +1 & 0 & --1 & --2 \\ \hline
\end{tabular}
\end{center}
This axial $U(1)_A$ symmetry restricts possible form of fermionic
mass terms and the most general terms compatible with it look like:
\begin{equation}
\frac{1}{m} {\cal L}_1 = - \frac{1}{2} \bar{\Psi}_\mu \sigma^{\mu\nu}
\Psi_\nu + i a_1 (\bar{\Psi}\gamma) \Omega + i a_2 (\bar{\Psi}\gamma)
\chi + a_3 \bar{\Omega}_i{}^j \Omega_j{}^i + a_4 \bar{\Omega} \Omega
+ a_5 \bar{\Omega} \chi + a_6 \bar{\rho} \lambda
\end{equation}
As for the vector fields, we have two complex doublet of scalars
which both could play the role of Goldstone fields. Straightforward
calculations show that it is the combination $z_i + \Phi_i$ that have
to be eaten by vector fields, leaving other combination $z_i -
\Phi_i$ as physical massive scalar fields. So we get mass terms for
bosonic fields:
\begin{equation}
{\cal L}_2 = \frac{m}{2\sqrt{2}} [ \varepsilon_{ij} C_\mu{}^i
\partial_\mu (\bar{z}^j + \bar{\Phi}^j) + (h.c.) ]
+ \frac{m^2}{2} C_\mu{}^i \bar{C}_{\mu i} - \frac{m^2}{4} |z_i -
\Phi_i|^2
\end{equation}
determine the coefficients for fermionic mass terms:
$$
a_1 = - a_2 = \frac{1}{\sqrt{2}}, \qquad a_3 = - a_4 = \frac{1}{2},
\qquad a_5 = a_6 = 1
$$
as well as structure of additional terms for fermionic
supertransformations:
\begin{eqnarray}
\frac{1}{m} \delta_1 \Psi_\mu &=& - C_\mu{}^i \eta_i -
\frac{i}{2\sqrt{2}} \gamma_\mu (z_i + \Phi_i) \varepsilon^{ij} \eta_j
 \qquad
\frac{1}{m} \delta_1 \rho = - \frac{1}{2} (\bar{z}^i - \bar{\Phi}^i)
\eta_i  \nonumber \\
\frac{1}{m} \delta_1 \Omega_i{}^j &=& - \frac{i}{\sqrt{2}} \gamma^\mu
[ C_\mu{}^i \eta_j - \delta_j{}^i C_\mu{}^k \eta_k ] + \frac{1}{2}
\varepsilon^{jk} (z_k - \Phi_k) \eta_i - \delta_i{}^j \Phi_k
\varepsilon^{kl} \eta_l   \\
\frac{1}{m} \delta_1 \chi &=& - \frac{i}{\sqrt{2}} \gamma^\mu
C_\mu{}^i \eta_i + z_i \varepsilon^{ij} \eta_j \qquad
\frac{1}{m} \delta_1 \lambda = - \frac{i}{\sqrt{2}} \gamma^\mu
\bar{C}_{\mu i} \varepsilon^{ij} \eta_j \nonumber
\end{eqnarray}
It is hardly comes as a surprise that besides global supersymmetry
and $U(2)$ symmetry our Lagrangian is invariant under the local gauge
transformations:
\begin{equation}
\delta \Psi_\mu = \partial_\mu \xi + \frac{im}{2} \gamma_\mu \xi,
\qquad \delta \Omega_i{}^j = \frac{m}{\sqrt{2}} \delta_i{}^j \xi,
\qquad \delta \chi = - \frac{m}{\sqrt{2}} \xi
\end{equation}
Making such transformation with:
$\xi = \frac{1}{\sqrt{2}} (z_i + \Phi_i) \varepsilon^{ij} \eta_j$
and introducing gauge invariant derivatives for scalar fields:
$$
D_\mu z_i = \partial_\mu z_i - \frac{m}{\sqrt{2}} \varepsilon_{ij}
C_\mu{}^j, \qquad D_\mu \Phi_i = \partial_\mu \Phi_i -
\frac{m}{\sqrt{2}} \varepsilon_{ij} C_\mu{}^j
$$
we obtain the final form of fermionic supertransformations:
\begin{eqnarray}
\delta \Psi_\mu &=& - \frac{i}{4} \sigma^{\alpha\beta} \gamma_\mu 
C_{\alpha\beta}{}^i \eta_i + \frac{1}{\sqrt{2}} D_\mu (z_i + \Phi_i)
\varepsilon^{ij} \eta_j \nonumber \\
\delta \rho &=& - \frac{1}{2\sqrt{2}} \sigma^{\alpha\beta}
\bar{C}_{\alpha\beta i}\varepsilon^{ij} \eta_j - \frac{m}{2}
(\bar{z}^i - \bar{\Phi}^i) \eta_i \nonumber  \\
\delta \Omega_i{}^j &=& - \frac{1}{2\sqrt{2}} \sigma^{\alpha\beta}
C_{\alpha\beta}{}^j \eta_i - i \varepsilon_{ik} \hat{D} z_j \eta^k +
\\
 && + \frac{m}{2} \varepsilon^{jk} (z_k - \Phi_k) \eta_i + \frac{m}{2}
\delta_i{}^j (z_k - \Phi_k) \varepsilon^{kl} \eta_l \nonumber \\
\delta \chi &=& i \hat{D} \Phi_i \varepsilon^{ij}  \eta_j + 
\frac{m}{2} (z_i - \Phi_i) \varepsilon^{ij} \eta_j \qquad
\delta \lambda = - i \hat{D} \bar{\Phi}^i \eta_i \nonumber
\end{eqnarray}

Such supermultiplet has to appear when $N=3$ or $N=4$ supergravity is
spontaneously broken up to $N=2$ and indeed such breaking turns out
to be possible as was shown in \cite{Zin86b,Zin87,TZ94,TZ94a} (see
also \cite{AFV02}).

\section{$N=3$ supermultiplet}

Our next example is massive $N=3$ supermultiplet containing one spin
3/2, six spin 1, fourteen spin 1/2 and fourteen spin 0 particles. It
easy to check that in the massless limit we will get one spin 3/2
supermultiplet $(3/2, 3 \otimes 1, 3 \otimes 1/2, 2 \otimes 0)$ and
three vector supermultiplets:
$$
\left( \begin{array}{c} \frac{3}{2} \\ 6 \otimes 1 \\ 14 \otimes
\frac{1}{2} \\ 14 \otimes 0 \end{array} \right) \quad \Rightarrow
\quad \left( \begin{array}{c} \frac{3}{2} \\ 3 \otimes 1 \\ 3 \otimes
\frac{1}{2} \\ 2 \otimes 0 \end{array} \right) \ \oplus \ 3 \otimes
\left( \begin{array}{c} 1 \\ 4 \otimes \frac{1}{2} \\ 6 \otimes 0
\end{array} \right)
$$
Really, this case is very similar to the previous one (and even
more simple due to the absence of hypermultiplet). Again it is crucial
that we have two triplets of (axial-)vector fields so we can arrange
them into one comples triplet. As a result we get $SU(3)$ invariant
supertransformations leaving the sum of kinetic terms invariant:
\begin{eqnarray}
\delta_0 \Psi_\mu &=& - \frac{i}{4} \sigma^{\alpha\beta} \gamma_\mu
C_{\alpha\beta}{}^ i \eta_i  \nonumber \\
\delta_0 \bar{C}_\mu{}^i &=& 2 (\bar{\Psi}_\mu \eta_i) 
+ i\sqrt{2} (\bar{\rho}_j{}^i \gamma_\mu \eta_j) \qquad 
\delta_0 C_{\mu i} = - i\sqrt{2} \varepsilon^{ijk}
(\bar{\chi}_j \gamma_\mu \eta_k) \nonumber \\
\delta_0 \chi^i &=& - \frac{1}{2\sqrt{2}} \varepsilon^{ijk}
\sigma^{\alpha\beta} \bar{C}_{\alpha\beta j} \eta_k - i \hat{\partial}
z \eta_i \qquad \delta_0 \bar{z} = 2 (\bar{\chi}^i \eta_i)  \\
\delta_0 \rho_i{}^j &=& - \frac{1}{2\sqrt{2}} \sigma^{\alpha\beta}
C_{\alpha\beta}{}^j \eta_i - i
\hat{\partial} \varepsilon^{ikl} \Phi_{jk} \eta_l \qquad
\delta_0 \lambda_i = - i \hat{\partial} \bar{\Phi}^{ij} \eta_j \qquad
\nonumber \\
\delta_0 \bar{\Phi}^{ij} &=& 2 (\bar{\rho}_k{}^i \varepsilon^{kjl}
\eta_l) \qquad \delta_0 \Phi_{ij} = 2 (\bar{\lambda}_i \eta_j)
\nonumber
\end{eqnarray}
Moreover, with the appropriate assignment of axial charges:
\begin{center}
\begin{tabular}{|c|c|c|c|c|c|c|} \hline
 field & $\eta_i$ & $\Psi_\mu$, $\rho_i{}^j$ & $\chi^i$ & $\lambda_i$
& $C_\mu{}^i$, $\Phi_{ij}$ & $z$ \\ \hline
$q_A$ & +1 & 0 & +2 & --2 & --1 & --3 \\ \hline
\end{tabular}
\end{center}
we gain $U(1)_A$ invariance as well. Among scalar fields there is
only one candidate for the role of Goldstone field, namely
antisymmetric part of $\Phi_{[ij]}$ leaving symmetric part as
physical massive scalars. So the mass terms for bosons look like:
\begin{equation}
{\cal L}_b = - \frac{m}{2\sqrt{2}} [ \varepsilon^{ijk} \bar{C}_{\mu i}
\partial_\mu \Phi_{jk} + h.c. ] + \frac{m^2}{2} [ (A_\mu{}^i)^2 +
(B_\mu{}^i)^2 - \bar{z} z - \bar{\Phi}^{(ij)} \Phi_{(ij)} ]
\end{equation}
while the most general fermionic mass terms compatible with $U(3)$
invariance have the form:
\begin{equation}
\frac{1}{m} {\cal L}_f = - \frac{1}{2} \bar{\Psi}_\mu \sigma^{\mu\nu}
\Psi_\nu + i a_1 (\bar{\Psi} \gamma) \rho_i{}^i + \frac{a_2}{2}
\bar{\rho}_i{}^j \rho_j{}^i + \frac{a_3}{2} \bar{\rho} \rho + a_4
\bar{\chi}^i \lambda_i
\end{equation}
Then the requirement that the whole Lagrangian be supersymmetric
fixes the unknown coefficients:
$$
a_1 = \frac{1}{\sqrt{2}}, \qquad a_2 = 1, \qquad a_3 = - 1, \qquad
a_4 = - 1
$$
which lead to invariance of the Lagrangian under the local gauge
transformations:
$$
\delta \Psi_\mu = \partial_\mu \xi + \frac{im}{2} \gamma_\mu \xi
\qquad \delta \rho_i{}^j = \frac{m}{\sqrt{2}} \delta_i{}^j \xi
$$
and also fixes the structure of fermionic supertransformations. By
using local gauge invariance and introducing gauge covariant
derivatives, supertransformations for fermions could be casted to the
form:
\begin{eqnarray}
\delta \Psi_\mu &=& - \frac{i}{4} \sigma^{\alpha\beta} \gamma_\mu
C_{\alpha\beta}{}^i \eta_i + \frac{i}{\sqrt{2}} \varepsilon^{ijk}
D_\mu z_{ij} \eta_k \nonumber \\
\delta \chi^i &=& - \frac{1}{2\sqrt{2}} \varepsilon^{ijk}
\sigma^{\alpha\beta} \bar{C}_{\alpha\beta j} \eta_k - i \hat{\partial}
z \eta_i + m \bar{z}^{(ij)} \eta_j \\
\delta \rho_i{}^j &=& - \frac{1}{2\sqrt{2}} \sigma^{\alpha\beta}
C_{\alpha\beta}{}^j \eta_i - i \gamma^\mu  \varepsilon^{ikl} (
\partial_\mu  z_{(jk)} + D_\mu z_{[jk]} ) \eta_l + m \varepsilon^{jkl}
z_{(ik)} \eta_l \nonumber \\
\delta \lambda_i &=& - i \gamma^\mu ( \partial_\mu  \bar{z}^{(ij)}
+ D_\mu \bar{z}^{[ij]} ) \eta_j + m z \eta_i \nonumber
\end{eqnarray}

Such massive supermultiplet really appears then $N=4$ supergravity is
broken up to $N=3$ \cite{Zin87,TZ94a,AFV02}. Note that there is an
interesting and still open question on the so called shadow
supermultiplets that appear in some compactifications \cite{FF01}. It
would be interesting to investigate possible interactions of such
massive $N=3$ supermultiplets with $N=3$ supergravity (without any
other supermultiplets).

\section{$N=2$ supermultiplet with central charge}

As is well known any massive supermultiplet without central charges
having $N \ge 4$ supersymmetries necessarily contains particles with
spin greater than 3/2. So we turn to the massive supermultiplets with
central charges and start with the simplest example --- with $N=2$
supersymmetry. This multiplet contains two equal sets of particles
corresponding to that of massive spin 3/2 supermultiplet with $N=1$
supersymmetry, so the counting of fields in the massless limit is the
same as before. But now we have to arrange all fields into massless
$N=2$ supermultiplets:
$$
2 \otimes \left( \begin{array}{c} \frac{3}{2} \\ 2 \otimes 1 \\
\frac{1}{2} \end{array} \right) \quad \Rightarrow \quad
2 \otimes \left( \begin{array}{c} \frac{3}{2} \\ 2 \otimes 1 \\
\frac{1}{2} \end{array} \right) \oplus \left( \begin{array}{c}
2 \otimes \frac{1}{2} \\ 4 \otimes 0 \end{array} \right)
$$
For the hypermultiplet we will use the same parametrisation as before:
\begin{eqnarray}
\delta \chi &=& - i \varepsilon^{ij} \hat{\partial} z_i \eta_j
\qquad \delta \bar{z}^i = 2 \varepsilon ^{ij} (\bar{\chi} \eta_j)
\nonumber \\
\delta \psi &=& - i \hat{\partial} \bar{z}^i \eta_i \qquad 
\delta z_i = 2 (\bar{\psi} \eta_i)
\end{eqnarray}
As for the spin 3/2 supermultiplets, the main trick is again to use
dual transformation for vector fields so that they enter through the
complex combinations only. Indeed, it is not hard to check that sum
of the kinetic terms for all fields is invariant under the following
global $N=2$ supertransformations:
\begin{eqnarray}
\delta \Psi_\mu &=& - \frac{i}{4} \sigma^{\alpha\beta} \gamma_\mu
C_{\alpha\beta i} \varepsilon^{ij} \eta^j \qquad
\delta \Omega_\mu = - \frac{i}{4} \sigma^{\alpha\beta} \gamma_\mu
\bar{C}_{\alpha\beta}{}^i \eta_i  \nonumber \\
\delta \bar{C}_\mu{}^i &=& 2 \varepsilon^{ij} (\bar{\Psi}_\mu \eta_j)
+ i\sqrt{2} \varepsilon^{ij} (\bar{\lambda} \gamma_\mu \eta_j) \qquad
\delta C_{\mu i} = 2 (\bar{\Omega}_\mu \eta_i)  + i\sqrt{2}
(\bar{\rho} \gamma_\mu \eta_i) \\
\delta \lambda &=& - \frac{1}{2\sqrt{2}} \sigma^{\alpha\beta}
C_{\alpha\beta i} \varepsilon^{ij} \eta_j \qquad \delta \rho = -
\frac{1}{2\sqrt{2}} \sigma^{\alpha\beta} \bar{C}_{\alpha\beta}{}^i
\eta_i \nonumber
\end{eqnarray}
The choice for the bosonic mass terms is unique:
\begin{equation}
{\cal L}_b = - \frac{m}{2} (\bar{C}_\mu{}^i \partial_\mu z_i +
h.c. ) + \frac{m^2}{2} \bar{C}_\mu{}^i C_{\mu i}
\end{equation}
This time we have only $SU(2) \simeq USp(2)$ global symmetry and no
axial $U(1)_A$ one so we have to consider the most general fermionic
mass terms:
\begin{eqnarray}
\frac{1}{m} {\cal L}_f &=& - \frac{1}{2} \bar{\Psi}_\mu
\sigma^{\mu\nu} \Psi_\nu + ia_1 (\bar{\Psi} \gamma) \lambda + i a_2
(\bar{\Psi} \gamma) \chi + a_3 \bar{\lambda} \chi + a_4 \bar{\chi}
\chi - \nonumber \\
 && - \frac{1}{2} \bar{\Omega}_\mu \sigma^{\mu\nu} \Omega_\nu + i b_1
(\bar{\Omega} \gamma) \rho + i b_2 (\bar{\Omega} \gamma) \psi + b_3
\bar{\rho} \psi + b_4 \bar{\psi} \psi
\end{eqnarray}
Indeed, the invariance of the total Lagrangian under the (corrected)
supertransformations could be achieved provided:
$$
a_1 = b_1 = \frac{1}{\sqrt{2}}, \qquad a_2 = b_2 = 1, \qquad
a_3 = b_3 = - \sqrt{2}, \qquad a_4 = b_4 = - \frac{1}{2}
$$
As in all previous cases, we have local gauge symmetries corresponding
to two spin 3/2 particles:
\begin{eqnarray}
\delta \Psi_\mu &=& \partial_\mu \xi_1 + \frac{im}{2} \gamma_\mu
\xi_1 \qquad \delta \lambda = \frac{m}{\sqrt{2}} \xi_1 \qquad
\delta \chi = m \xi_1 \nonumber \\
\delta \Omega_\mu &=& \partial_\mu \xi_2 + \frac{im}{2} \gamma_\mu
\xi_2 \qquad \delta \rho = \frac{m}{\sqrt{2}} \xi_2 \qquad
\delta \psi = m \xi_2
\end{eqnarray}
In this, with the help of these transformations, introducing gauge
invariant derivative $ D_\mu z_i = \partial_\mu z_i -m C_{\mu i}$ we
obtain final form of fermionic supertransformations:
\begin{eqnarray}
\delta \Psi_\mu &=& - \frac{i}{4} \sigma^{\alpha\beta} \gamma_\mu
C_{\alpha\beta i} \varepsilon^{ij} \eta_j + D_\mu z_i \varepsilon^{ij}
\eta_j \nonumber \\
\delta \Omega_\mu &=& - \frac{i}{4} \sigma^{\alpha\beta} \gamma_\mu
\bar{C}_{\alpha\beta}{}^i \eta_i + D_\mu \bar{z}^i \eta_i \nonumber \\
\delta \lambda &=& - \frac{1}{2\sqrt{2}} \sigma^{\alpha\beta}
C_{\alpha\beta i} \varepsilon^{ij} \eta_j \qquad
\delta \chi = - i \hat{D} z_i \varepsilon^{ij} \eta_j \\
\delta \rho &=& - \frac{1}{2\sqrt{2}} \sigma^{\alpha\beta}
\bar{C}_{\alpha\beta}{}^i \eta_i \qquad \delta \psi = - i \hat{D}
\bar{z}^i \eta_i \nonumber
\end{eqnarray}

Such supermultiplet could appear when $N=4$ supergravity is broken up
to $N=2$ and two massive gravitini have equal masses
\cite{Zin87,TZ94a,AFV02}.

\section{$N=4$ supermultiplet with central charge}

Our last example --- massive $N=4$ supermultiplet with central
charge. Such multiplets contains twice as many fields as massive $N=2$
supermultiplet without central charge and in the massless limit it
gives just two massless spin 3/2 supermultiplets:
$$
2 \otimes \left( \begin{array}{c} \frac{3}{2} \\ 4 \otimes 1 \\
6 \otimes \frac{1}{2} \\ 4 \otimes 0 \end{array} \right) \quad
\Rightarrow \quad 2 \otimes \left( \begin{array}{c} \frac{3}{2} \\ 4
\otimes 1 \\ 7 \otimes \frac{1}{2} \\ 8 \otimes 0 \end{array} \right)
$$
This time even using the usual trick with vector fields it is
impossible to obtain complete $SU(4)$ symmetry. Indeed \cite{AAFL02}
maximum symmetry that we can get here is the $USp(4)$ one. Thus we
introduce $USp(4)$ invariant antisummetric tensor $\omega_{[ij]}$ such
that $\omega^{ij} \omega_{jk} = - \delta^i{}_k$ and use it to
construct $USp(4)$ invariant form of supertransformations for massless
supermultiplets:
\begin{eqnarray}
\delta \Psi_\mu &=& - \frac{i}{4} \sigma^{\alpha\beta}
\gamma_\mu C_{\alpha\beta i} \omega^{ij} \eta_j \qquad
\delta \Omega_\mu = - \frac{i}{4} \sigma^{\alpha\beta}
\gamma_\mu \bar{C}_{\alpha\beta}{}^i \eta_i \nonumber \\
\delta \bar{C}_\mu{}^i &=& 2 (\bar{\Psi}_\mu \omega^{ij} \eta_j)
 + 2 i (\bar{\rho}^{ij} \gamma_\mu \eta_j) \qquad
\delta C_\mu{}^i =  2 (\bar{\Omega}_\mu \eta_i) + 2 i
(\bar{\lambda}_{ij} \gamma_\mu \omega^{jk} \eta_k) \nonumber \\
\delta \lambda^{ij} &=& - \frac{1}{2} \sigma^{\alpha\beta}
\bar{C}_{\alpha\beta}{}^{[i} \omega^{j]k} \eta_k - i \sqrt{2}
\hat{\partial} z_{[i} \eta_{j]} + \frac{i}{\sqrt{2}} \omega^{ij}
 \hat{\partial} z_k \omega^{kl} \eta_l \qquad
\delta \chi = - i \hat{\partial} \bar{z}^i \eta_i \\
\delta \rho_{ij} &=& - \frac{1}{2} \sigma^{\alpha\beta}
C_{\alpha\beta[i} \eta_{j]} - i \sqrt{2} \hat{\partial}
\bar{\Phi}^{[i} \omega^{j]k} \eta_k - \frac{i}{\sqrt{2}} \omega_{ij}
\hat{\partial} \bar{\Phi}^k \eta_k \qquad
\delta \psi = - i \hat{\partial} \Phi_i \omega^{ij} \eta_j \nonumber
\\
\delta \bar{z}^i &=& 2 \sqrt{2} (\bar{\lambda}^{ij} \eta_j) -
\sqrt{2} (\bar{\lambda}^{kl} \omega_{kl} \omega^{ij} \eta_j) \qquad
\delta z_i = 2 (\bar{\chi} \eta_i) \nonumber \\
\delta \Phi_i &=& 2 \sqrt{2} (\bar{\rho}_{ij} \omega^{jk}
\eta_k) + \sqrt{2} (\bar{\rho}_{kl} \omega^{kl} \eta_i) \qquad
\delta \bar{\Phi}^i = 2 (\bar{\psi} \omega^{ij} \eta_j) \nonumber
\end{eqnarray}
Then subsequent calculations lead us to the following mass terms for
bosons:
\begin{equation}
{\cal L}_b = \frac{m}{2\sqrt{2}} C_{\mu i} \omega^{ij} 
\partial_\mu (z_j - \Phi_j) + h.c. + \frac{m^2}{2} \bar{C}_\mu{}^i
C_{\mu i} - \frac{m^2}{4} (\bar{z}^i + \bar{\Phi}^i) (z_i + \Phi_i)
\end{equation}
from which we see that combination $z_i - \Phi_i$ plays the role of
Goldtone fields while $z_i + \Phi_i$ remains as physical massive
scalars. As for the fermionic fields, their mass terms turn out to be:
\begin{eqnarray}
\frac{1}{m} {\cal L} &=& - \frac{1}{2} \bar{\Psi}_\mu
\sigma^{\mu\nu} \Psi_\nu + \frac{i}{2} (\bar{\Psi} \gamma) \rho -
\frac{i}{\sqrt{2}} (\bar{\Psi} \gamma) \chi  + \frac{1}{2}
\omega^{ik} \omega^{jl} \bar{\rho}_{ij} \rho_{kl} - \frac{1}{4}
\bar{\rho} \rho + \frac{1}{\sqrt{2}} \bar{\rho} \chi - \nonumber \\
 && - \frac{1}{2} \bar{\Omega}_\mu \sigma^{\mu\nu} \Omega_\nu -
\frac{i}{2} (\bar{\Omega} \gamma) \lambda -
\frac{i}{\sqrt{2}}(\bar{\Omega}\gamma) \psi +
\frac{1}{2} \omega_{ik} \omega_{jl} \bar{\lambda}^{ij}
\lambda^{kl} - \frac{1}{4} \bar{\lambda} \lambda
- \frac{1}{\sqrt{2}} \bar{\lambda} \psi 
\end{eqnarray}
which corresponds to invariance under the following two local gauge
transformations:
\begin{eqnarray*}
\delta \Psi_\mu &=& \partial_\mu \xi_1 + \frac{im}{2} \gamma_\mu
\xi_1, \qquad \delta \rho_{ij} = \frac{m}{2} \omega_{ij} \xi_1, \qquad
\delta \chi = - \frac{m}{\sqrt{2}} \xi_1 \\
\delta \Omega_\mu &=& \partial_\mu \xi_2 + \frac{im}{2} \gamma_\mu
\xi_2, \qquad \delta \lambda^{ij} = - \frac{m}{2} \omega^{ij} \xi_2,
\qquad \delta \psi = - \frac{m}{\sqrt{2}} \xi_2
\end{eqnarray*}
With the help of these transformations and introducing gauge
invariant objects:
$$
D_\mu z_i = \partial_\mu z_i - \frac{m}{\sqrt{2}} \omega_{ij}
\bar{C}_\mu{}^j, \qquad
D_\mu \Phi_i = \partial_\mu \Phi_i + \frac{m}{\sqrt{2}} \omega_{ij}
\bar{C}_\mu{}^j
$$
we obtain final form of fermionic supertransformations:
\begin{eqnarray*}
\delta \Psi_\mu &=& - \frac{i}{4} \sigma^{\alpha\beta}
\gamma_\mu C_{\alpha\beta i} \omega^{ij} \eta_j - \frac{1}{\sqrt{2}}
D_\mu (\bar{z}^i - \bar{\Phi}^i) \eta_i \\
\delta \Omega_\mu &=& - \frac{i}{4} \sigma^{\alpha\beta}
\gamma_\mu \bar{C}_{\alpha\beta}{}^i \eta_i  + \frac{1}{\sqrt{2}}
D_\mu (z_i - \Phi_i) \omega^{ij} \eta_j \\
\delta \lambda^{ij} &=& - \frac{1}{2} \sigma^{\alpha\beta}
\bar{C}_{\alpha\beta}{}^{[i} \omega^{j]k} \eta_k - i \sqrt{2}
\hat{D} z_{[i} \eta_{j]} + \frac{i}{\sqrt{2}} \omega^{ij}
 \hat{D} z_k \omega^{kl} \eta_l - \\
 &&  - \frac{m}{\sqrt{2}} [ (z_k + \Phi_k) \omega^{k[i} \omega^{j]l}
\eta_l + \frac{1}{2} \omega^{ij} (z_k + \Phi_k) \omega^{kl}  \eta_l ]
\\
\delta \rho_{ij} &=& - \frac{1}{2} \sigma^{\alpha\beta}
C_{\alpha\beta[i} \eta_{j]} - i \sqrt{2} \hat{D} \bar{\Phi}^{[i}
\omega^{j]k} \eta_k - \frac{i}{\sqrt{2}} \omega_{ij} \hat{D}
\bar{\Phi}^k \eta_k + \\
 && + \frac{m}{\sqrt{2}} [ (\bar{z}^k + \bar{\Phi}^k) \omega_{k[i}
\eta_{j]} + \omega_{ij} (\bar{z}^k + \bar{\Phi}^k) \eta_k ] \\
\delta \chi &=& - i \hat{D} \bar{z}^i \eta_i  + \frac{m}{2} (\bar{z}^i
+ \bar{\Phi}^i) \eta_i  \\
\delta \psi &=& - i \hat{D} \Phi_i \omega^{ij} \eta_j  + \frac{m}{2}
(z_i + \Phi_i) \omega^{ij} \eta_j
\end{eqnarray*}

\section*{Conclusion}

Thus we give explicit construction of massive spin 3/2 supermultiplets
out of the massless ones and this gives us important and model
 independent information om the structure of supergravity models
where such supermultiplets could arise as a result of spontaneous
supersymmetry breaking. Also we hope that experience gained will be
helpful in investigation of massive supermultiplets with arbitrary
superspins.

\end{document}